\documentclass{aastex}          %% The manuscript based on AASTeX v5.x
\usepackage{spr-astr-addons}    %% mimicing ASTR journal style
\begin{document}
%% Article title
%
  \title{Correlation analysis of radio properties and accretion-disk luminosity for low luminosity AGNs}
%% Running heads

\shortauthors{Su, Liu \& Zhang}

\author{Renzhi Su\altaffilmark{1,3}}
\and
\author{Xiang Liu\altaffilmark{1,2}}
\and
\author{Zhen Zhang\altaffilmark{1}}
%\author{Xiaolong Yang\altaffilmark{1,3}}
%\and
%\author{Zhenhua Han\altaffilmark{4}}

%\email{surenzhi14@mails.ucas.ac.cn}

\altaffiltext{1}{Xinjiang Astronomical Observatory, Chinese
Academy of Sciences, 150 Science 1-Street, Urumqi 830011, PR
China}

\altaffiltext{2}{Key Laboratory of Radio Astronomy, Chinese
Academy of Sciences, Urumqi 830011, PR China}

\altaffiltext{3}{Graduate University of Chinese Academy of
Sciences, Beijing 100049, PR China}

%\altaffiltext{4}{Xinjiang Normal University, Urumqi, 830054, PR China}

\begin{abstract}

The correlation between the jet power and accretion disk luminosity is investigated and analyzed with our model for 7 samples of low luminosity active galactic nuclei (LLAGNs). The main results are: (1) the power-law correlation index ($P_{jet} \propto L_{disk}^{\mu}$) typically ranges $\mu=0.4-0.7$ for the LLAGN samples, and there is a hint of steep index for the LLAGN sample which hosted by a high fraction of elliptical galaxies, and there are no significant correlation between the $\mu$ and the LLAGN types (Seyfert, LINER); (2) for $\mu \approx$1, as noted in Liu et al., the accretion disk dominates the jet power and the black hole (BH) spin is not important, for the LLAGN samples studied in this paper we find that the $\mu$ is significantly less than unity, implying that BH spin may play a significant role in the jet power of LLAGNs; (3) the BH spin-jet power is negatively correlated with the BH mass in our model, which means a high spin-jet efficiency in the `low' BH-mass LLAGNs; (4) an anti-correlation between radio loudness and disk luminosity is found, which is apparently due to the flatter power-law index in the jet-disk correlation of the LLAGNs, and the radio loudness can be higher in the LLAGNs than in luminous AGNs/quasars when the BH spin-jet power is comparable to or dominate over the accretion-jet power in the LLAGNs. The high radio-core dominance of the LLAGNs is also discussed.

\end{abstract}


\section{Introduction}

Early efforts were made to investigate the properties of low luminosity or `dwarf' active galactic nuclei in nearby galaxies in Elvis, Soltan \& Keel (1984), Filippenko \& Sargent (1985), Keel (1985), Koratkar et al. (1995), and Ho et al. (1997a). From optical spectroscopic survey of the nuclear regions ($<$200 pc) of 486 nearby (typical distance of $\sim$18 Mpc) galaxies (a nearly complete,
magnitude-limited sample of galaxies with $B_{T}<12.5 mag$ and declination greater than $0^{\circ}$) using the Hale 5 m telescope at
Palomar Observatory, Ho et al. (1997b) obtained about 200 nuclei show spectroscopic evidence of active galactic nuclei consisting of Seyfert nuclei, low-ionization nuclear emission-line regions (LINERs, see Heckman et al. 1980), and transition
objects (composite LINER/H II-nucleus systems), including 46 nuclei that show broad $H_{\alpha}$ emission. Operationally, Ho et al. (1997a) define
`low-luminosity' or `dwarf' AGNs to be those with
$L_{H\alpha}<10^{40}$ erg/s (over 85\% of the Palomar AGNs lie below this value), and designed to select
objects on the faint end of the luminosity function with much lower luminosities than `classical' Seyfert nuclei and quasars. The host galaxies of these low luminosity AGNs (LLAGNs) show either early type (Hubble type E and S0) or late type spiral galaxies. The scaling of bolometric luminosity of accretion disk to the emission line luminosity is $L_{disk}=2000L_{H\alpha}$ in Netzer (2009) for luminous AGNs, and a much smaller factor of 300 is suggested for low luminosity AGNs (Ho 2009). The LLAGN is often defined with the optical line luminosity or with nuclear X-ray luminosity typically $<10^{42}$erg/s (e.g. Ho 2009), which are 1-3 orders of magnitude smaller than in classical Seyfert galaxies (Terashima et al. 2002). The bolometric luminosity can be as high as $\sim10^{44}$erg/s in some LLAGN samples. However, the Eddington accretion rate (Eddington ratio) is much lower (typically around $10^{-5}$) in the LLAGNs than that in luminous AGNs and quasars (usually $\ge 10^{-2}$).

The LLAGNs have shown different properties from luminous AGNs and quasars, with smaller black hole masses accreting at low rates. The host galaxies of LLAGNs are mostly lack of classical (elliptical-galaxy-like) bulges, so the classical $M_{bh}-\sigma$ relation may be not necessary for the LLAGNs and the coevolution from central black hole to outskirt matter of LLAGN is weak (Kormendy \& Ho 2013). But the nuclear parameters like the disk luminosity, radio core/jets and hard X-ray emission may still be correlated.

There is a much higher detection rate ($>$50\%) of radio loud nuclei in the LLAGNs than that ($\sim15\%$) in the luminous AGNs (Ho \& Ulvestad 2001; Nagar et al. 2002). The radio morphologies of the radio active LLAGNs show predominately a core feature, only a small fraction of them show long-jet features or extended emission on scale greater than a kpc (Ho \& Ulvestad 2001; Nagar et al. 2002). The radio spectra of the cores are mostly flat, but strong relativistic jet or beaming is rare. These striking radio properties may be suggestive of that the jet/outflow production mechanism of LLAGNs is different from luminous AGNs. Liu \& Han (2014) and Liu et al. (2016) studied the correlation indices between radio jet power and disk luminosity for different type of AGNs, including a small sample of LLAGNs, and found that the FRII sources have a strong linear correlation between jet power and disk luminosity, whereas the LLAGNs may have a flatter power-law correlation index between jet power and disk luminosity, which might be due to possible contribution of jet power from black hole spin (Blandford \& Znajek 1977). In this paper, we revisit the radio properties of LLAGNs from several larger LLAGN samples and analyze mainly the correlation between jet power and accretion disk for the LLAGNs, in order to test whether or not the power-law index of the correlation is indeed more flat for the LLAGNs and to study how the black hole spin could affect the radio properties of LLAGNs.

\section[]{Correlation between radio power and disk luminosity in LLAGNs}

We analyze the correlation between jet power and disk luminosity of LLAGN samples from the literature in this paper. The bolometric luminosity of accretion disk is thought to be proportional to the optical line luminosity $H_{\alpha}$, O[III], etc, with different scaling factors for different optical lines, e.g. with $L_{disk}=300L_{H\alpha}$ for the LLAGNs as suggested by Ho (2009). The disk luminosity of LLAGN can also be estimated with the hard X-ray luminosity which thought to come from the nuclear region: $L_{bol}=15.8 L_{X}$ (erg/s) in 2-10 keV (Ho 2009) is used in our analysis.

The radio jet power $P_{j}$ can be estimated with the radio luminosity at 1.4 GHz ($L_{1.4}[erg/s]$) and which is derived from the relation $P_{j}=2.32\times10^{20} (f/3)^{3/2} (P_{1.4})^{6/7}[W/Hz] (erg/s)$ (the factor $f$ is a constant within range of 1-20, which $f=10$ we used does not affect our analysis in the following) in Liu \& Han (2014) as the form $P_{j}=2.05\times10^{7}(L_{1.4})^{6/7}$(erg/s), in which the work done on the jet environment by jet kinetic energy  has been considered (Willott et al. 1999; Cavagnolo et al. 2010).

The correlation model of the jet power and disk luminosity is $P_{j} \propto L_{disk}^{\mu}$ as proposed in Liu \& Han (2014), it can be written as
\begin{equation}
log (P_{j}) = \mu \times log (L_{disk}) + const.
\end{equation}

A linear correlation (i.e the $\mu\approx1$ in the equation 1) between jet power and disk luminosity has been found in FR II type quasars (van Velzen \& Falcke 2013) and PG quasars (Liu \& Han 2014), which is explained with the accretion-dominated jet in Liu et al. (2016). On the other hand, a flatter correlation index between radio jet power and disk luminosity has been found in both the narrow line galaxies and black hole X-ray binaries, which implies the black hole (BH) spin may play an important role in the jet power (Liu et al, 2016). The BH spin-produced jet (`spin-jet' for short) can be derived from Tchekhovskoy et al. (2011), Yuan \& Narayan (2014) and Liu et al. (2016) as:
\begin{equation}
P_{spin}\propto a^{2}m^{3\kappa-2}\dot m^{\kappa}/\varepsilon \propto a^{2}m^{2(\kappa-1)}L_{disk}^{\kappa}/\varepsilon,
\end{equation}

$a$ is the dimensionless spin parameter, m is black hole mass in solar mass unit, $\dot m$ is mass accretion rate in Eddington accretion rate unit, $\varepsilon$ is the disk radiative efficiency. The $\Phi \propto (\Phi_{MAD})^{\kappa}$ with $\kappa<1$ is defined in Liu et al. (2016), $\Phi$ is magnetic flux and the $\Phi_{MAD}$ is the upper limit of $\Phi$, i.e. $\Phi_{MAD}$: the saturation flux of the magnetically arrested disk (MAD) (Tchekhovskoy et al. 2011; Yuan \& Narayan 2014). The system which have not reached the MAD limit have been referred to as SANE
(¡°standard and normal evolution¡±, Narayan et al. 2012; Sadowski et al. 2013), in which $\Phi$ spans from a small value to a magnetic flux just
below $\Phi_{MAD}$ for a spinning BH, and this accords with $\kappa<1$ in our case.

We note that the BH spin-jet power has a power-law index $\mu=\kappa<1$ versus disk luminosity in the equation 2. The spin-jet power is inversely proportional to the disk radiative efficiency $\varepsilon$, and has a negative correlation with the BH mass since the $\kappa-1$ is always negative. The result is interesting that smaller black holes tend to have higher spin-jet efficiency. If we, for example, take a median value of $\kappa=0.5$, the equation 2 will have a simple form of $P_{spin}\propto a^{2}m^{-1}L_{disk}^{0.5}/\varepsilon$.

\subsection[]{LLAGN samples and model-fitting results}

We analyze 7 LLAGN samples searched from the literature in the following. The correlation analysis between jet power and disk luminosity was not done before for most of the LLAGN samples, we re-analyze the samples with adding more radio data by searching from the literature when they are not available from the samples, and the accretion disk luminosity can be estimated from optical line luminosity or hard X-ray luminosity.

1) The first LLAGN sample is defined with the $H_{\alpha}$ line luminosity of $<10^{40}$erg/s and width (FWHM) $>1000$ km/s by Ho et al. (1997b) from the Palomar AGN survey, which consists of 46 nuclei, of them 22 are Seyfert nuclei, 24 are LINERs including 2 transition objects, and the broad $H_{\alpha}$ emission indicates the origin of the emission from AGN rather from starburst nuclei. We searched from the literature (mostly the NVSS data) for the flux density at 1.4 GHz of the broad-line LLAGNs, and estimated their jet power assuming that the radio emission is from the AGN, and the accretion disk luminosity is estimated with the $H_{\alpha}$ line luminosity (Ho et al. 1997b) with the relation of $L_{disk}=300L_{H\alpha}$. With the linear regression analysis of equation 1, the power-law index is obtained, and the sample shows a tight correlation between the radio jet power and disk luminosity, see the result listed in Table~\ref{tab1}.

2) The LLAGN sample selected by Ho \& Ulvestad (2001) from the Palomar catalog consists of 52 Seyfert Nuclei, with 30 type 2
and 22 type 1 objects. The radio morphologies of the LLAGNs have been found to be predominantly of a compact core, either unresolved or slightly resolved, occasionally accompanied by elongated, jetlike features (Ho \& Ulvestad 2001). Sources in this sample are all Seyfert nuclei, and  $\sim$40\% of the sources also appear in the `broad-line' LLAGN sample above. The Seyfert 1 galaxies are found to have somewhat stronger radio sources than Seyfert 2 galaxies (Ulvestad \& Ho 2001). With the radio data at 1.4 GHz searched by us and the disk luminosity estimated with $H_{\alpha}$, we made a linear regression analysis and obtained the power-law index between jet power and disk luminosity and the correlation coefficient as listed in Table~\ref{tab1}.

3) A sample of 53 LLAGNs have been complied by Jang et al. (2014) defined with the bolometric luminosity $<10^{42}$ erg/s and with dynamical BH mass measured and good-quality X-ray data available. Of them 43 are from the Palomar AGN sample. With the radio data at 1.4 GHz searched by us and the disk luminosity estimated with the X-ray data, a linear regression analysis is made with the equation 1 and the result is listed in Table~\ref{tab1}. The power-law correlation index is quite steep (0.75) which is probably due to the higher fraction (43\%) of elliptical hosts in the sample than in other samples.

4) Nagar et al. (2002) have studied the nearest LLAGNs ($<19$ Mpc) from the Palomar AGN sample. With radio data at 1.4 GHz searched by us and disk luminosity estimated with $H_{\alpha}$ line luminosity, a linear regression analysis is made and the result is in Table~\ref{tab1}.

5) Nagar et al. (2005) compiled all ($\sim$200) LLAGNs from the Palomar Sample. We collected radio data available at 1.4 GHz, and disk luminosity is estimated with $H_{\alpha}$ line luminosity, and a linear regression analysis is made for the equation 1 and the result is shown in Table~\ref{tab1}.

\begin{figure}
  \includegraphics[width=8cm]{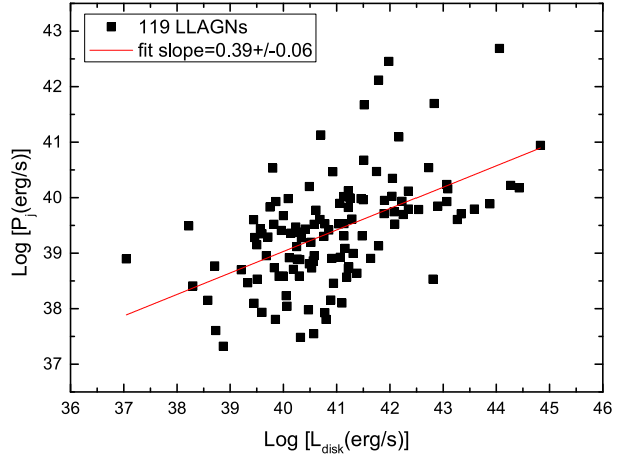}
  \caption{$logP_{jet}$ vs. $log L_{disk}$ ($L_{disk}=15.8L_{X(2-10keV)}$) for the Ho (2009) sample, and a linear fit has been made.}
  \label{fig1}
\end{figure}

6) Ho (2009) compiled a sample of 175 LLAGNs from the Palomar Sample with collected nuclear X-ray luminosities of the LLAGNs from the literature, most of the sources also appear in the Nagar et al (2005) sample. We searched radio data at 1.4 GHz and a linear regression analysis is made, the result is shown in Fig.~\ref{fig1} and in Table~\ref{tab1}.

\begin{figure}
  \includegraphics[width=8cm]{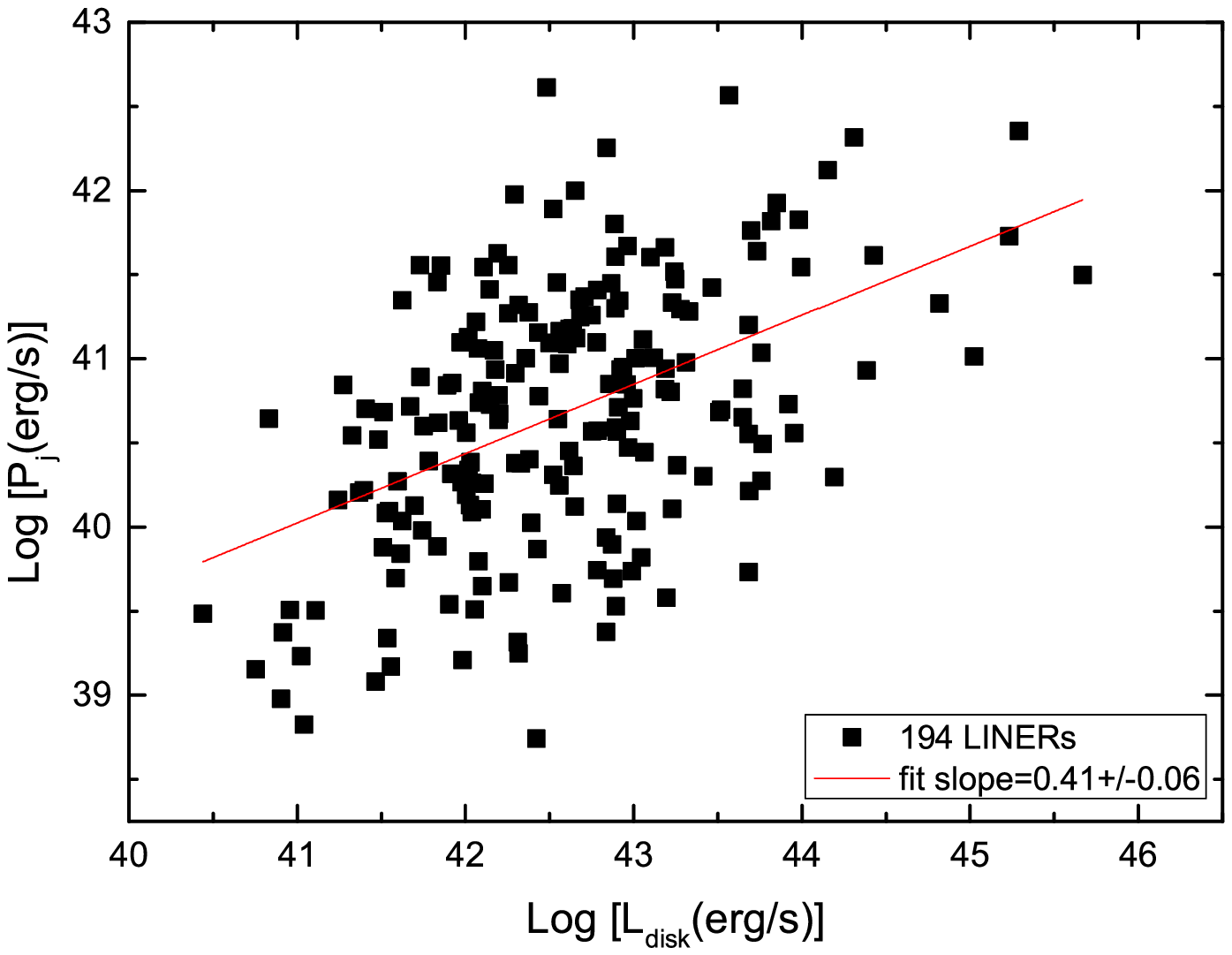}
  \caption{$logP_{jet}$ vs. $log L_{disk}$ ($L_{disk}=15.8L_{X(2-10keV)}$) for the LINERs with flux densities $>$0.48 mJy at 1.4 GHz in Nisbet \& Best (2016) sample, and a linear fit has been made.}
  \label{fig2}
\end{figure}

7) Nisbet and Best (2016) compiled a large sample of LINERs constructed by cross-match data from the 3XMM-DR4
and SDSS-DR7 catalogues with redshift $z<0.3$, and with AGN black hole mass, 1.4 GHz flux density and hard X-ray (2-10 keV) data available. They have fitted for the correlation between radio luminosity and X-ray luminosity with all data including the upper limits of radio flux density. We reanalyze the linear regression fit with only the reliable flux densities ($>$0.48 mJy at 1.4 GHz) of 194 LINERs, the result is shown in Fig.~\ref{fig2} and in Table~\ref{tab1}. The power-law correlation index of 0.41 in the LINERs is close to the result of Ho (2009) which consists of low luminosity Seyferts, LINERs and transition objects. This LLAGN sample contains LINERs only which distributes in much larger distances than the Ho (2009) sample.

To summarize the results, we find that the jet power is significantly correlated with the disk luminosity for all the samples, see the correlation coefficient and the null probability in Table~\ref{tab1}. The sample 1 to 6 are actually sub-samples of the complete Palomar AGN sample which stated in the introduction, except the sample 3 (Jang sample) in which 19\% of sources not from the Palomar survey. As mentioned, there is a steeper $\mu$ that possibly caused by the high fraction of elliptical hosts in the Jang sample (see Table~\ref{tab1}), as we know that the FRII quasars have shown steep indices $\mu\approx1$ in van Velzen \& Falcke (2013) and Willott et al. (1999) which hosted mostly by elliptical galaxies. We find no significant correlations between the LLAGN types (Sy, L) and the index $\mu$ in Table~\ref{tab1}. Moreover, the median distances of sources in the sample 1-6 are in the narrow range of 13.7-25.5 Mpc in Table~\ref{tab1}, the sample 7 i.e. the Nisbet and Best (2016) sample has much larger distance (median redshift of 0.104) than the Palomar sample, however, the resulted $\mu$ is similar in Table~\ref{tab1} e.g. for the Ho (2009) and Nisbet-Best (2016) samples.

We notice that (1) the sample sizes in Table~\ref{tab1} are smaller than the original sample sizes because some sources have no radio data available from the literature, and (2) the detection of compact radio core or hard X-ray emission is a signature of AGN rather than an HII nuclei for the transition objects which were not well identified optically in some of the samples we analyzed.

\begin{table*}

 \caption[]{The statistical information, correlation strength, and the power-law correlation index fitted for the LLAGN samples analyzed in this paper. The columns are the sample name (Br$H_{\alpha}$LLAGN: the broad $H_{\alpha}$ line LLAGNs in Ho et al. (1997b), Ho-Ulvestad: the Seyfert type LLAGNs in Ho \& Ulvestad (2001), Jang: the LLAGNs in Jang et al. (2014), Nagar (2002): the LLAGNs in Nagar et al. (2002), Nagar (2005): the LLAGNs in Nagar et al. (2005), Ho (2009): the Palomar LLAGNs in Ho (2009), Nisbet-Best: the LINERs with radio flux density higher than 0.48 mJy at 1.4 GHz in Nisbet \& Best (2016), respectively); number of sources; fraction in percent of different Hubble type of host galaxy (E: elliptical galaxy, S0, and S: spiral galaxy); fraction in percent of different AGN class (Sy: Seyfert galaxy, L: LINER, U: unclassified LLAGN, and T: the transition object); the largest (median distance in brackets) luminosity distance in Mpc unit; Pearson correlation coefficient; probability of non-correlation; and the power-law index ($\mu$) in jet power vs. disk luminosity or the $\rho$ (minus value) in radio loudness vs. disk luminosity.}
         $$
         \begin{tabular}{cccccccc}
\hline
  \hline
    \noalign{\smallskip}
    1&2 &3 &4 & 5&6 &7 & 8 \\
 Sample & N & Hubble type (\%) & AGN class (\%) &  $D_{L}$  & Coef. & $P_{null}$ & $\mu$ or $\rho$\\
    & &  & & &  &
    &   \\

\hline
 \noalign{\smallskip}
%-----------------------------------------------------------------------------------------------------------------------
  Br$H_{\alpha}$LLAGN & 44  &E(16),S0(27),S(57)   & Sy(50),L(45),T(5)  &91.6(18.3)&0.68 & 3.0E-7 & 0.69$\pm$0.11   \\
%-----------------------------------------------------------------------------------------------------------------------
  Ho-Ulvestad &47    & E(6),S0(30),S(64)   &Sy(100)             & 70.1(20.4)  &  0.62   & 2.5E-6   & 0.56$\pm$0.10   \\
%-----------------------------------------------------------------------------------------------------------------------
  Jang        &44     &E(43),S0(32),S(25)    &Sy(36),L(32),U(32) &99.3(15.9)  &0.74     &9.7E-8   &0.75$\pm$0.12  \\
%-----------------------------------------------------------------------------------------------------------------------
  Nagar(2002) & 79    & E(9),S0(23),S(68)   & Sy(29),L(35),T(36)   &  18.9(16.5) &  0.39  & 1.8E-3  &0.39$\pm$0.12   \\
%-----------------------------------------------------------------------------------------------------------------------
  Nagar(2005) & 151   & E(13),S0(26),S(61) &Sy(26),L(42),T(32)  &70.6(18.2)  & 0.51   & 1.8E-11   & 0.54$\pm$0.07   \\
%-----------------------------------------------------------------------------------------------------------------------
  Ho(2009)    & 119   &E(18),S0(27),S(55)  & Sy(33),L(44),T(23)  & 91.6(17.0) &  0.54   & 2.0E-10 & 0.39$\pm$0.06  \\
 (R vs. Ldisk) &              &                    &                    &    &-0.62           & 7.2E-14  & -0.55$\pm$0.06 \\
%-----------------------------------------------------------------------------------------------------------------------
  Nisbet-Best  & 194         &                             &  L(100)             &  (z=0.104)    &  0.47   & 2.9E-12 &  0.41$\pm$0.06   \\
  (R vs. Ldisk) &              &                    &                    &    & -0.50   & 6.4E-14   & -0.52$\pm$0.06        \\

           \noalign{\smallskip}
            \hline
           \end{tabular}{}
        % \end{array}
         $$
         \label{tab1}
   \end{table*}

\subsection[]{Radio loudness and radio core dominance}

The radio loudness is often defined with the ratio of radio luminosity over optical line luminosity or over hard X-ray luminosity $R=L_{1.4}/L_{X}$ (Terashima \& Wilson 2003). Because $L_{disk}=15.8 L_{X}$, $P_{jet}\propto L_{X}^{\mu}$, and $P_{jet}\propto L_{1.4}^{6/7}$, for the radio loudness to X-ray (or to optical line which has the same form, see Liu \& Han 2014), we have
\begin{equation}
R=L_{1.4}/L_{X}=L_{X}^{(7/6)\mu-1}=L_{X}^{\rho} \propto L_{disk}^{\rho},
\end{equation}
where $\rho=(7/6)\mu-1$.

\begin{figure}
  \includegraphics[width=8cm]{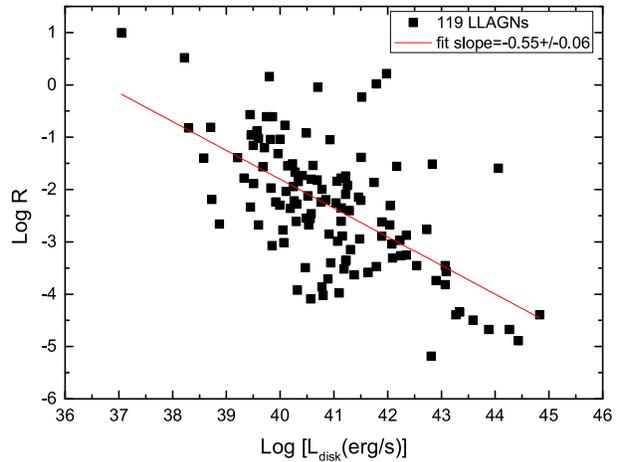}
  \caption{$logR(radio loudness£º R=L_{1.4}/L_{X})$ vs. $log L_{disk}$ ($L_{disk}=15.8L_{X(2-10keV)}$) for the Ho (2009) sample, and a linear fit has been made.}
  \label{fig3}
\end{figure}

\begin{figure}
  \includegraphics[width=8cm]{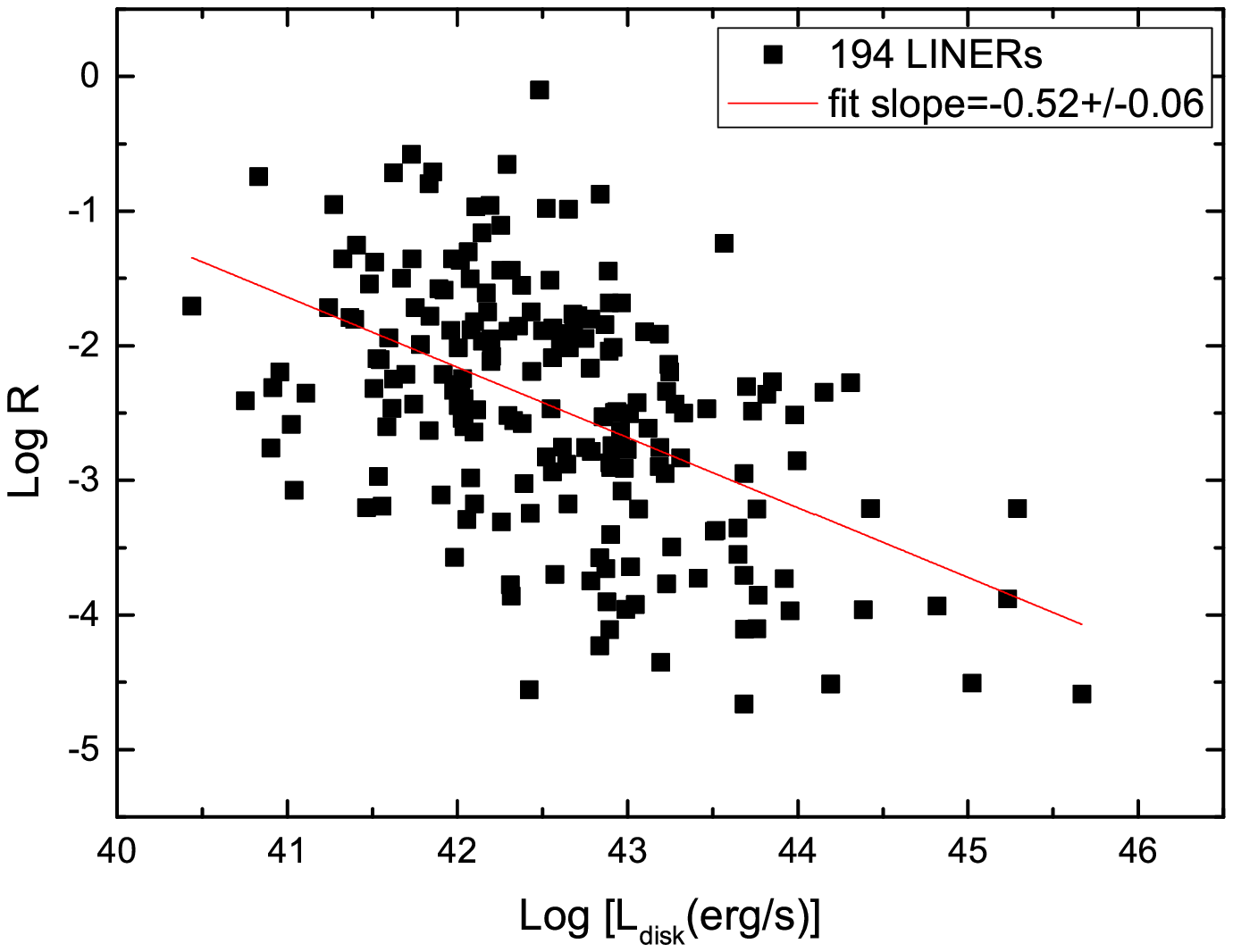}
  \caption{$logR(radio loudness£º R=L_{1.4}/L_{X})$ vs. $log L_{disk}$ ($L_{disk}=15.8L_{X(2-10keV)}$) for the LINERs with flux densities $>$0.48 mJy at 1.4 GHz in Nisbet \& Best (2016) sample, and a linear fit has been made.}
  \label{fig4}
\end{figure}

We calculate the radio loudness $R=L_{1.4}/L_{X}$ and disk luminosity $L_{disk}=15.8 L_{X}$, and carried out a model fit with the equation 3, for all the samples. The results show an anti-correlation (i.e. $\rho<0$) for all the samples. Here we present the results only for the Ho (2009) sample and Nisbet-Best (2016) sample in Fig.~\ref{fig3} and Fig.~\ref{fig4} and in Table~\ref{tab1}. This anti-correlation is expected in the equation 3 when $\rho=(7/6)\mu-1<0$, i.e. $\mu<6/7\simeq0.86$. In the samples of LLAGNs we studied, $\mu$ is in range of 0.39-0.75 (Table~\ref{tab1}), less than 0.86, so that lead to the anti-correlation between the radio loudness and accretion disk luminosity. The anti-correlation also apparently explains the high fraction of radio-loud LLAGNs observed, i.e. the lower the disk luminosity (or accretion rate) the higher the radio loudness of the LLAGNs.

\begin{figure}
  \includegraphics[width=8cm]{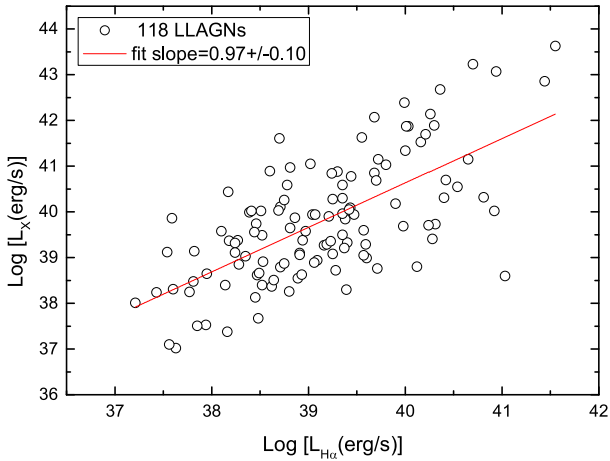}
  \caption{Hard X-ray luminosity $logL_{X(2-10keV)}$ versus $H_{\alpha}$ line luminosity $logL_{H_{\alpha}}$ for the Ho (2009) sample, and a linear fit has been made.}
  \label{fig5}
\end{figure}

We also calculated the radio loudness with the $H_{\alpha}$ line luminosity $R_{opt}=L_{1.4}/L_{H_{\alpha}}$ for the Ho (2009) sample, and find that the $R_{opt}$ is also anti-correlated with the disk luminosity which estimated from the hard X-ray luminosity, but with a larger fitting error compared to the $-0.55\pm0.06$ in Fig.~\ref{fig3}. The larger scatter of the anti-correlation between $R_{opt}$ and the X-ray estimated disk power could be due to the larger uncertainty of the $H_{\alpha}$ fluxes (typical error of 30\%-50\%, Ho 2009), and Ho (2009) suggested that the hard X-ray data are more accurate and better to be used to estimate accretion disk luminosity than the $H_{\alpha}$ line luminosity, although there is a positive correlation between the $H_{\alpha}$ line luminosity and the hard X-ray luminosity as seen in Fig.~\ref{fig5}.

The LLAGNs are predominantly to show a compact core, occasionally accompanied by short jetlike features or diffuse emission (Ho \& Ulvestad 2001). The fraction of core-dominated sources is $\sim$80\% in the Ho \& Ulvestad (2001) sample, and it is $>$90\% in Nagar et al. (2005) sample (defined with the ratio of peak flux density over total flux density $>$50\%). This high radio-core dominance is about 2 times greater in fraction than that of 40\% (defined with the core flux density $>$50\% of total flux density) in normal radio-loud galaxies and quasars of the complete Caltech-Jodrell Bank VLBI Survey sample (Polatidis et al. 1995).

\section[]{Summary and discussion}

We have carried out the model fit to almost all known LLAGN samples searched from the literature, and find the following results focused on the LLAGNs, in which for the BH spin-dominated jet we discuss in detail the BH spin-jet efficiency, the BH-spin effects on the radio loudness and radio core dominance of LLAGNs, which are not stressed in our previous papers (Liu \& Han 2014; Liu et al. 2016).

(1) The power-law correlation index between jet power and disk luminosity is significantly less than unity (typically $\mu=0.4-0.7$) for most of the samples. There is a hint of steep index for the LLAGN sample which hosted by a high fraction of elliptical galaxies, and there are no significant correlation between the $\mu$ and the LLAGN types (Seyfert, LINER).

(2) According to our model this flatter index of the LLAGNs could be due to a significant contribution of jet power from black hole spin, because it accords with the weaker dependence on the disk luminosity in equation 2, i.e. the averaged $\bar{\mu}\sim0.5$ from Table~\ref{tab1} (excluding the Jang sample) and so $\kappa=\bar{\mu}\sim0.5$ in the spin-jet power of equation 2.

(3) The BH spin may not be extreme for most of the LLAGNs, the spin-jet power could be still significant due to its negative correlation with the BH mass for the `low' BH-mass LLAGNs, i.e. the spin-jet efficiency is roughly $\eta_{spin}\propto a^{2}m^{-1}$ for $\kappa=\bar{\mu}\sim0.5$, which means a high spin-jet efficiency in the `low' BH-mass LLAGNs. On the other hand, the jet power in luminous AGNs and quasars are usually dominated by disk accretion because of the linear correlation found between the jet power and disk luminosity (van Velzen \& Falcke 2013; Liu et al. 2016), in which the BH spin-jet power would be significantly reduced for super-massive ($>10^{8} M_{\odot}$) BHs by the $\eta_{spin}\propto a^{2}m^{2(k-1)}$ and the $(k-1)<0$ in equation 2. The BH spin of AGNs is difficult to measure, only a few have been estimated (Reynolds 2014) which are statistically not enough to test our model, but we will try to do this in the future.

(4) An anti-correlation between radio loudness and disk luminosity is found, which is apparently due to the flatter power-law index in the jet-disk correlation of the LLAGNs. To discuss the radio loudness further, assuming the accretion-jet luminosity $L_{acc}$, the spin-jet luminosity $L_{spin}$, and an optical line luminosity e.g. $L_{H_{\alpha}}$, the radio loudness can be defined as $R_{acc}=L_{acc}/L_{H_{\alpha}}$ for a pure accretion-jet, and $R_{tot}=(L_{acc}+L_{spin})/L_{H_{\alpha}}$ for total jet power consisting of accretion-jet and spin-jet, then we have $R_{tot}>R_{acc}$. This suggests that the radio loudness can be higher in the LLAGNs than in luminous AGNs/quasars when the BH spin-jet power is comparable to or dominate over the accretion-jet power in the LLAGNs, because that as mentioned the jet power in luminous AGNs/quasars is dominated by the accretion disk (i.e. $L_{acc} \gg L_{spin}$) due to their linear jet-disk correlation.

Moreover, we discuss that the radio jets in the LLAGNs are often core-dominated with occasionally short jets or extended emission, this high core dominance is probably due to that the jets are still very young, so that they have no enough time to grow larger. However, we are actually not sure if the LLAGNs are really young radio sources. Alternatively, in our model mentioned above, the BH spin-jet in the LLAGNs could be more compact and steady than the accretion-jet because the spin energy is tapped from the ergosphere of spinning black hole (Penrose 1969) i.e immediate vicinity of the BH, by contrast the accretion-jet is usually formed in a much larger scale. But careful simulations are required for this hypothesis to be tested. We note that the beaming effect should not be responsible for the high radio-core dominance of LLAGNs, because the jets in the LLAGNs are mostly not relativistic and small viewing angles are rare in a complete LLAGN sample such as the Palomar sample.

\section*{Acknowledgments}

 Helpful comments from the reviewer are appreciated that have improved the paper. This research was supported from the following funds: the programme of the Light in China's Western Region (grant no. XBBS201324); the 973 Program 2015CB857100; the Key Laboratory of Radio Astronomy, Chinese Academy of Sciences; and the National Natural Science Foundation of China (No.11273050).

\end{document}